\documentclass[]{aa} 
\usepackage{graphicx}
\usepackage{natbib}
\bibpunct{(}{)}{;}{a}{}{,} 
\newcommand{\TO}[0]{{-}}

\newcommand{\micron}{\mbox{$\mu$m}}

\newcommand{\CIIw}{\ion{C}{ii}}
\newcommand{\CII}{[\ion{C}{ii}]}
\newcommand{\CIw}{\ion{C}{i}}
\newcommand{\CI}{[\ion{C}{i}]}

\newcommand{\OI}{[\ion{O}{i}]}

\newcommand{\OIII}{[\ion{O}{iii}]}

\newcommand{\HII}{\ion{H}{ii}}
\newcommand{\HI}{\ion{H}{i}}

\newcommand{\NII}{[\ion{N}{ii}]}
\newcommand{\lsun}{L$_{\odot}$}
\newcommand{\msun}{M$_{\odot}$}

\begin{document}

\title{Photon dominated regions in the spiral arms of M83 and M51}
  
   \author{C.\,Kramer\inst{1} \and
          B.\,Mookerjea\inst{1} \and
          E.\,Bayet\inst{2} \and
          S.\,Garcia-Burillo\inst{3} \and
          M.\,Gerin\inst{2} \and
          F.P.\,Israel\inst{4} \and
          J.\,Stutzki\inst{1} \and 
          J.G.A.\,Wouterloot\inst{5} 
          }


   \institute{KOSMA, I. Physikalisches Institut,
              Universit\"at zu K\"oln,
              Z\"ulpicher Stra\ss{}e 77,
              50937 K\"oln, Germany \and
%
              Radioastronomie Millimetrique: UMR 8540 du CNRS,
              Laboratoire de Physique de l'ENS, 24 Rue Lhomond, 
              75231 Paris cedex 05, France \and
              Centro Astronomico de Yebes,
              IGN, E-19080 Guadalajara, Spain \and
              Sterrewacht Leiden,
              P.O. Box 9513,
              2300 RA Leiden, The Netherlands \and
              Joint Astronomy Centre,
              660 N. A'ohoku Place, 96720 Hilo, HI, USA 
}
   \offprints{C.\,Kramer, \email{kramer@ph1.uni-koeln.de}}
   \date{Received 3 May 2005 / Accepted }
   
   \abstract{We present \CI\ $^3$P$_1$--$^3$P$_0$ spectra at four
     spiral arm positions and the nuclei of the nearby galaxies M83
     and M51 obtained at the JCMT. The spiral arm positions lie at
     galacto-centric distances of between 2\,kpc and 6\,kpc. This data
     is complemented with maps of CO 1--0, 2--1, and 3--2, and ISO/LWS
     far-infrared data of \CII\ (158\,$\mu$m), \OI\ (63\,$\mu$m), and
     \NII\ (122\,$\mu$m) allowing for the investigation of a complete
     set of all major gas cooling lines. From the intensity of the
     \NII\ line, we estimate that between 15\% and $30$\% of the
     observed \CII\ emission originate from the dense ionized phase of the
     ISM. The analysis indicates that emission from the diffuse
     ionized medium is negligible.  In combination with the FIR dust
     continuum, we find gas heating efficiencies below $\sim0.21\%$ in
     the nuclei, and between 0.25 and 0.36\% at the outer positions.
     Comparison with models of photon-dominated regions (PDRs) of
     Kaufman et al.  (1999) with the standard ratios
     \OI(63)/\CII$_{\rm PDR}$ and (\OI(63)$+$\CII$_{\rm PDR}$) vs.
     TIR, the total infrared intensity, yields two solutions.  The
     physically most plausible solution exhibits slightly lower
     densities and higher FUV fields than found when using a full set
     of line ratios, \CII$_{\rm PDR}$/\CI(1--0), \CI(1--0)/CO(3--2),
     CO(3--2)/CO(1--0), \CII/CO(3--2), and, \OI(63)/\CII$_{\rm PDR}$.
     The best fits to the latter ratios yield densities of
     $10^4$\,cm$^{-3}$ and FUV fields of $\sim\,G_0=$20--30 times the
     average interstellar field without much variation. At the outer
     positions, the observed total infrared intensities are in perfect
     agreement with the derived best fitting FUV intensities. The
     ratio of the two intensities lies at 4--5 at the nuclei,
     indicating the presence of other mechanisms heating the dust.
     \CI\ area filling factors lie below 2\% at all positions,
     consistent with low volume filling factors of the emitting gas.
     The fit of the model to the line ratios improves significantly
     if we assume that \CI\ stems from a larger region than CO
     2--1. 
%
%
     Improved modelling would need to address the filling factors of
     the various submm and FIR tracers, taking into consideration the
     presence of density gradients of the emitting gas by including
     cloud mass and size distributions within the beam.
%
     \keywords{Galaxies: ISM, Galaxies: structure, Galaxies:
       individual: M83, M51, ISM: Structure, Infrared: Galaxies,
       Submillimeter} }

   \maketitle
%

\section{Introduction} 

Neutral atomic carbon is thought to form predominantly in surface
layers of molecular clouds where \CIIw\ recombines and CO is
dissociated due to the far-UV photons governing the chemical
reactions.  FUV photons (6\,eV$<h\nu<13.6$\,eV) are primarily
responsible for the heating of the surface regions via photoelectric
effect on dust grains while at larger depths cosmic-ray induced
heating will dominate. 
These regions are referred to as photo dissociation regions or, more
generally, as photon dominated regions {(PDRs)}
\citep[][]{tielens1985,stoerzer1996,kaufman1999}.  PDR models
take into account the relevant physical processes, solve
simultaneously for the chemistry (using an extensive chemical network)
and the thermal balance, as a function of cloud depth.
%
%
It is found that the ratio of \CII/\CI\ is an accurate tracer of the
FUV field \citep{gerin_phillips2000}, parametrized by $G_0$ in units
of the Habing-field $1.6\,10^{-3}$\,erg s$^{-1}$cm$^{-2}$
\citep{habing1968}. Another important parameter governing the depth at
which \CIw\ forms is the ratio of density over FUV field $n/G_0$
\citep[][]{tielens1985}. This ratio also determines the efficiency of
converting FUV photons to gas heating, i.e. the photoelectric heating
efficiency $\epsilon$ \citep{bakes_tielens1994}.

While the Milky Way survey of FIR lines conducted with COBE/FIRAS
\citep{fixsen1999} showed that \CII\ is the dominant cooling line, it
also showed the importance of the two finestructure lines of \CI.
Both lines are ubiquitous and the two lines together amount to 75\% of
the total cooling of all rotational CO lines in the inner galaxy. This
picture has also emerged from extragalactic observations of the \CI\ 
1--0 line. These show again that the cooling due to \CIw\ and CO are
of the same order of magnitude for most galaxies \citep{bayet2004,
  israel_baas2003, israel_baas2002, israel_baas2001}.
\CIw\ is found to be a good tracer of molecular gas, possibly more
reliable than CO \citep{gerin_phillips2000}.



Several coordinated mapping studies of nearby galaxies have been
started during the past years. The BIMA SONG survey \citep{regan2001}
has aimed at obtaining the $^{12}$CO emission of 1--0 and 2--1
rotational lines at high spatial resolutions. There exist
velocity-integrated \CII\ observations of large samples of galaxies
with the KAO \citep{stacey1991} and with ISO \citep{malhotra2001},
hereafter MKH01, and \citep{leech1999,negishi2001} at $\sim1'$
resolution.
The SINGS Spitzer Legacy Project \citep{kennicutt2003} has started
imaging 75 galaxies in the infrared,
 including M51.
In the coming years, both SOFIA and the Herschel Space Observatory are
expected to provide velocity-resolved \CII\ data at resolutions of
$\sim10''$, complementary to many current single dish observations of
CO and \CI.

In external galaxies where a large number of clouds or even GMCs fill
the beam it is difficult to seperate the different contributions and
judge their importance.  A long standing problem is that a substantial
fraction of the \CII\ emission may originate from the diffuse ionized
and neutral medium. Comparison with \NII\ helps to estimate the
fraction originating from PDRs, but usually with large uncertainties
due to the varying chemical and excitation conditions in different
galactic environments \citep[e.g.][]{contursi2002}. 
%
The present study is 
part of the preparatory work for future airborne and space missions
like SOFIA and Herschel. In addition, it may serve as template for
studies of e.g.  \CI\ and CO in high-$z$ galaxies which have recently
become possible
\citep{weiss2005,walter2004,pety2004,neri2003,weiss2003}.

Most extragalactic observations of atomic carbon have so far
concentrated on the bright galactic nuclei or enhanced emission of
edge-on galaxies.
Here, we compare observations of the two nuclei of M83 and M51 with
pointed observations at spiral arm positions which show enhanced star
forming activity. The galacto-centric distances of the selected outer
positions lie between 1.8 and 5.8\,kpc.
%
We combine observations of atomic carbon with low and mid-$J$ CO and
$^{13}$CO data, as well as FIR \CII, \OI($63\,\mu$m), and
\NII($122\,\mu$m)\ data from the ISO data-base and thus include the
brightest gas cooling lines of the far-infrared and submillimeter
regime.

\begin{center}
\begin{table}[h*]
\caption[]{\label{tab-the-sample}
{\small Basic properties of M83 and M51.
$D_{25}\times d_{25}$ is the optical diameter from the RC3 catalogue.
$L_{\rm FIR}$ is the surface integrated far-infrared luminosity between 42.5 and 122.5$\,\mu$m
using the listed distance.
$F_{60}/F_{100}$ is the IRAS color ratio of the surface integrated flux densities 
at $60\,\mu$m and $100\,\mu$m, corrected for extinction. 
%
%
References: 
(1) RC3 catalogue \citep{devaucouleurs1991}, 
(2) \citet{talbot1979,tully1988},
(3) \citet{rice1988},
(4) \citet{sandage_tammann1975},
(5) \citet{devaucouleurs1979}
}}
\begin{tabular}{lrrrrr}
\hline \hline
                             & M83           & M51 \\ 
\noalign{\smallskip} \hline \noalign{\smallskip}
RA(2000)                     &    13:37:00.5 & 13:29:52.7 \\
DEC(2000)                    & $-29$:51:55.3 & 47:11:43   \\
%
%
Type                         & SAB(s)c$^{(1)}$   & SA(s)bc pec$^{(1)}$ \\
Distance [Mpc]               & 3.7$^{(5)}$         & 9.6 $^{(4)}$ \\
$10''$ correspond to         & 179\,pc             & 465\,pc \\
Heliocentric velocity [kms$^{-1}$]  & 516$^{(1)}$ & 463$^{(1)}$ \\ 
Position Angle [deg]         & 45 $^{(2)}$        & 170 \\ 
Inclination [deg]            & 24 $^{(2)}$         & $20$ \\
$D_{25}\times d_{25}$ [$'$] &  $12.9\times11.5$$^{(1)}$ & $11.2\times6.9$$^{(1)}$ \\
$L_{\rm FIR}$ [$10^9$\,\lsun]& $7.1^{(3)}$         & $14^{(3)}$ \\ 
$F_{60}/F_{100}$             & $0.43^{(3)}$        & $0.44^{(3)}$ \\ 
$F_{60}$ [Jy]                & $286^{(3)}$         & $85^{(3)}$ \\ 
\noalign{\smallskip} \hline \noalign{\smallskip}
\end{tabular}
\end{table}
\end{center}

\subsection{M83}

M83 (NGC\,5236) is the most nearby CO-rich grand-design spiral galaxy,
seen almost face-on (Table\,\ref{tab-the-sample}). It has a pronounced
bar, with two well-defined spiral arms connected to the starburst
nucleus.
In this paper, we adopt a distance of 3.7\,Mpc
\citep{devaucouleurs1991} though recent observations of Cepheids
indicate a slightly larger distance of 4.5\,Mpc \citep{thim2003}.



Low-$J$ CO maps were obtained by
\citet{crosthwaite2002,lundgren2004,lundgren2004_2,dumke2001,sakamoto2004}.
\CI\ observations of the center were conducted by
\citet{israel_baas2002} and \citet{petitpas_wilson1998}.
Pointed KAO observations report strong FIR fine-structure lines
towards the nucleus with a rapid fall-off towards the arms
\citep{crawford1985}.  
%

Here, we present new \CI\ data of the center and two spiral arm
positions on the north-eastern arm and south-western bar-spiral
transition zone.  The emission of \CII, \NII(122), and \OIII(88)
observed with ISO/LWS (Brauher 2005, priv. com.) is strongly
enhanced in these interface regions indicating greatly enhanced star
formation rates. ISO/LWS emission from the center was analyzed by
\citet{negishi2001}. The {\bf north-eastern arm} was previously
studied by \citet{lord_kenney1991} and \citet{rand1999} who presented
OVRO interferometric $^{12}$CO 1--0 maps. The eastern position at
($89'',38''$) presented here corresponds to the bright feature \#6
compiled by \citet[][Table\,4]{rand1999}.  About $15''$ to the east of
the CO arm newly formed stars form the optical arm and an HI ridge.
At \#6, the CO and dust arms coincide while they are offset further to
the south.
%
%
Position ($-80'',-72''$) studied here corresponds to a CO 1--0 peak in
the {\bf south-western bar-spiral transition zone} which exhibits a massive
GMC complex and luminous \HII\ regions \citep{kenney_lord1991}.  Note
that only less than 5\% of the single dish flux is recovered by the
interferometric maps \citep{rand1999}.  Thus, relatively smoothly
distributed diffuse molecular gas is completely missed.

%

\subsection{M51}

The nearby grand-design spiral galaxy M51 (NGC\,5194) seen almost face
on (Table\,\ref{tab-the-sample}) is interacting with its small
companion NGC\,5195, which lies $4.5'$ to the north. M51 is a Seyfert 2
galaxy \citep{ho1997}. The central AGN is surrounded by a
$\sim100$\,pc disk \citep{kohno1996} of dense and warm gas
\citep{matsushita1998}.

A large amount of observational data are available for this object,
including an extended KAO map of \CII\ \citep{nikola2001}.
\citet{garnett2004} used ISO/LWS data of the M51 \HII\ region CCM\,10,
to study the ionized gas, finding that abundances are roughly solar.
%
%
%

Several single-dish studies mapped the low-lying rotational $^{12}$CO
and $^{13}$CO transitions. CO 1--0 and 2--1 was mapped by
\citet{gb1993a,gb1993b} and \citet{nakai1994,tosaki2002}. In this
grand-design spiral, CO is tightly confined to the spiral arms.  Maps
of CO 3--2 and 4--3 were obtained at the HHT by
\citet{nieten1999,wielebinski1999,dumke2001} who show that warm
molecular gas is extended in M51 at galacto-centric distances of at
least upto $100''$, resp. 5\,kpc.  Single-dish observations of neutral
carbon were so far obtained only in the center region by
\citet{gerin_phillips2000,israel_baas2002}, and Israel, Tilanus, Baas,
2005, in prep.).

Aperture synthesis maps were obtained by
\citet{aalto1999,sakamoto1999} and \citet{regan2001} in CO 1--0 at
resolutions of $4''-6''$.  Recently, \citet{matsushita2004} mapped the
inner region in $^{12}$CO 3--2. 

Here, we selected 
two positions at the spiral arms lying in the northeastern and the
southwestern zones, i.e. at $72'',84''$ and $-84'',-84''$, of enhanced
\CII\ emission tracing enhanced star formation \citep{nikola2001}.
ISO/LWS data are available for these positions, and for the center
\citet{negishi2001} and Brauher (2005, priv. com.). The \HII\ region
studied by \citet{garnett2004} using ISO/LWS, CCM\,10, lies at about
$+148''$,$+45''$ \citep{carranza1969}, $1.5'$ to the north-west of
$72''$,$84''$.



\begin{center}
\begin{table}[h*]
\caption[]{\label{tab-data-sets}
{\small Overview of the \CI, CO, and $^{13}$CO data of M83 and M51. 
$\theta_{\rm b}$ is the telescope half power beamwidth.
All spectra are scaled to main beam brightness temperatures, 
$T_{\rm mb}=T_{\rm A}^*/\eta_{\rm{mb}}$. 
References: 
1: new data for this paper; 
2: \citet{crosthwaite2002};
3: Bayet et al. (priv. com.);
4: \citet{wielebinski1999,nieten1999,dumke2001};
5: \citet{gb1993a}. 
%
The last column indicates whether single pointings or maps are available.
The mapped CO data was gauss-smoothed to the $80''$ ISO/LWS resolution.
}}
\begin{tabular}{rrrrrcl}
\hline \hline
& Line & $\theta_{\rm{b}}$ & Telescope & $\eta_{\rm{mb}}$ & Refs. \\
&      & ($''$) \\
\hline
\multicolumn{7}{l}{{\bf M83} at (0,0),(89,38),($-80,-72$)} \\ 
            & \CI\ 1--0 & 10 & JCMT 15m & 0.52 & 1 & point. \\
            & CO 1--0   & 21 & IRAM 30m & 0.78 & 1 & point. \\
            & CO 1--0   & 55 & NRAO 12m & 0.88 & 2 & map \\ 
            & CO 2--1   & 10 & IRAM 30m & 0.57 & 1 & point. \\
            & CO 2--1   & 28 & NRAO 12m & 0.56 & 2 & map \\ 
            & CO 3--2   & 25 & CSO 10m  & 0.75 & 3 & map \\ 
            & $^{13}$CO 1--0 & 22 & IRAM 30m & 0.79 & 1 & point. \\
            & $^{13}$CO 2--1 & 10 & IRAM 30m & 0.60 & 1 & point. \\ 
\hline 
\multicolumn{7}{l}{{\bf M51} at (0,0), (72,84), ($-84,-84$)} \\
                          & \CI\ 1--0 & 10 & JCMT 15m & 0.52 & 1 & point \\ 
                          & CO 1--0   & 21 & IRAM 30m & 0.65 & 5 & map \\ 
%
                          & CO 2--1   & 10 & IRAM 30m & 0.46 & 5 & map \\ 
                          & CO 3--2   & 22 & HHT 10m  & 0.50 & 4 & map \\ 
                          & $^{13}$CO 1--0 
                                      & 22 & IRAM 30m & 0.79 & 1 & point. \\ 
                          & $^{13}$CO 2--1 
                                      & 10 & IRAM 30m & 0.60 & 1 & point. \\ 
\hline
\end{tabular}
\end{table}
\end{center}

\begin{figure}[htb]
\centering
\includegraphics[angle=0,height=10cm]{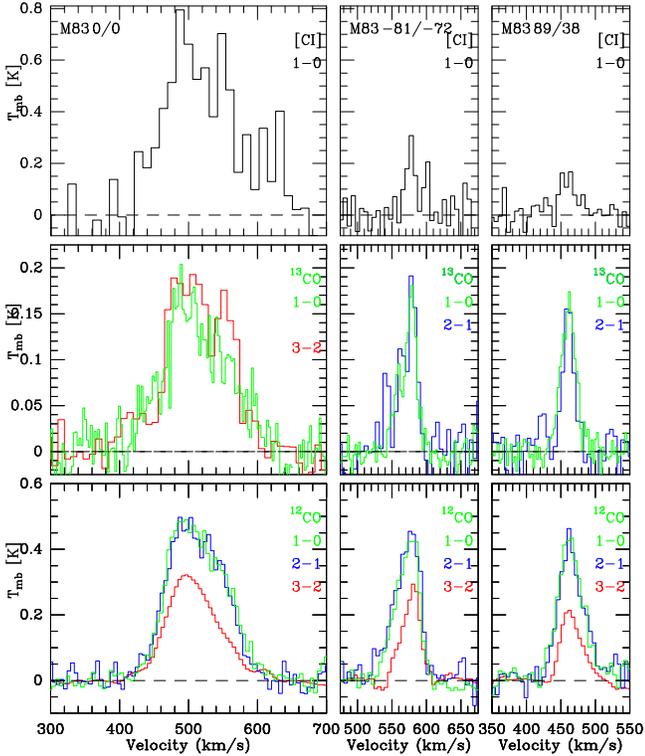} 
\caption{Spectra of M83 at the central and two spiral arm positions 
  (Table\,\ref{tab-data-sets}). Offsets are given in arcseconds
  relative to the (0,0) position (Table\,\ref{tab-the-sample}).
  Velocities are relative to LSR. All spectra are on the $T_{\rm mb}$
  scale. The $^{12}$CO data are at a common resolution of $80''$. The
  \CI\ and $^{13}$CO data are at their original resolutions listed in
  Table\,\ref{tab-data-sets}.
\label{fig-m83-spec}
}
\end{figure}

\begin{figure}[htb]
\centering
\includegraphics[angle=0,height=10cm]{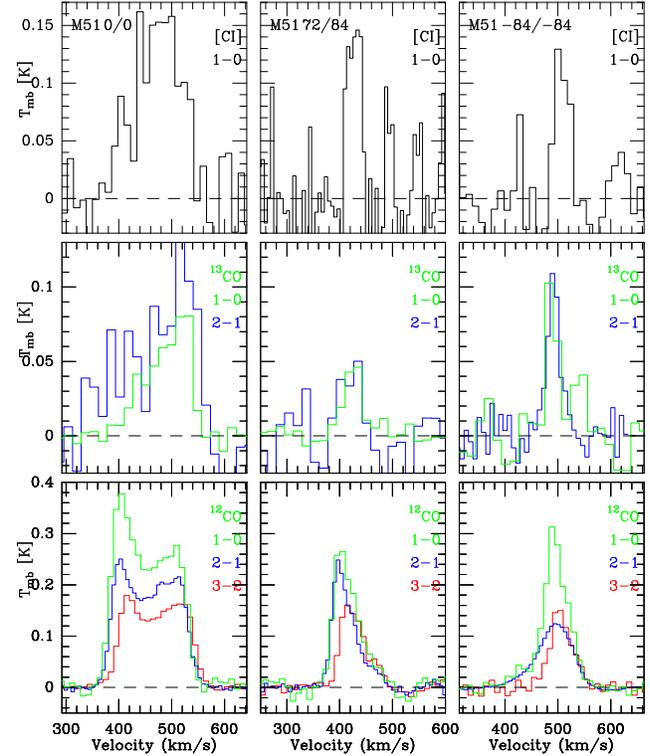}
\caption{Spectra of M51  at the central and two spiral arm positions 
  (cf. Fig.\,\ref{fig-m83-spec}).
\label{fig-m51-spec}
}
\end{figure}



\section{Data sets} 

We present here observations of \CI, CO, and $^{13}$CO spectra at
four spiral arm positions and the centers of M51 and M83
(Table\,\ref{tab-data-sets}).  We combine these with ISO/LWS FIR
spectral line data at all six positions together with the FIR
continuum derived from HIRES/IRAS 60$\,\mu$m and 100$\,\mu$m maps.

\subsection{\CI\ data taken at the JCMT}

We have observed the fine structure transition of atomic carbon (\CI)
at 492\,GHz (609~$\mu$m, $^3$P$_1\TO^3$P$_0$; hereafter $1\TO0$)
in M51 (3 positions) and M83 (3 positions) using the JCMT 15m
telescope.  We used receiver RxW with a single mixer and the DAS
autocorrelator.  Observations were carried out during
 35 hours in May and June 2003.
We used the double-beamswitch observing mode with a wobbler throw of
$\pm3'$ in the direction of the major axis,
i.e.  in the direction of the largest velocity gradient.
Pointing was checked using SCUBA after an initial alignment with RxW
and Jupiter at the start of each shift. It was found to be accurate to
within $2-3''$. The atmospheric zenith opacity at 225\,GHz
varied slowly between 0.1 and 0.05, corresponding to a $\tau$ of 
2 and 1 at 492\,GHz. 
After merging the DAS autocorrelator spectra using the SPECX software,
further data analysis was done using the CLASS/GILDAS package of IRAM.

\subsection{CO data taken at the IRAM 30m MRT}

We have observed the $^{12}$CO and $^{13}$CO 1--0 and 2--1 rotational
transitions at all 
six positions in M51 and M83 using the IRAM 30m telescope.  These
observations were carried out in double beam switch mode with a
wobbler throw of $\pm4'$ 
using the filterbank of 1\,MHz resolution for the 3\,mm band and the
4\,MHz filterbank for the 1\,mm band.  Observations were carried out
during 15 hours on July, 23rd and 26th, and on September, 10th, 2004.
Pointing and focus were checked and corrected every $\sim2$ hours. The
pointing accuracy was better than $4''$. The amount of precipitable
water vapour varied slowly between 10 and $\sim20$\,mm. Telescope
parameters are listed in Table\,\ref{tab-data-sets}.

%
%
%

\begin{table}
\begin{center}
\caption{\label{tab-isodata} FIR continuum and line intensities
  in units of $10^{-6}$\,erg\,s$^{-1}$\,cm$^{-2}$\,sr$^{-1}$.
   The absolute calibration error is assumed to be 15\% \citep{gry2003}.
%
%
}
\begin{tabular}[h]{crrrr}
\hline \hline
Positions                     & HIRES       & \multicolumn{3}{l}{ISO/LWS} \\
{$(\Delta\alpha,\Delta\delta$)} & FIR   & \CII      &  \NII     & \OI      \\ 
{$('','')$}                   &     & $158\mu$m & $122\mu$m & $63\mu$m     \\ 
\hline
 {\bf M83:} \\   
 (0,0) &     32\,$10^3$ &    81.8 &    14.1 &    86.7 \\  
 (-80,-72) &     7.2\,$10^3$ &    38.6 &    5.3 &    27.3 \\  
 (89,38) &     6.6\,$10^3$ &    33.4 &    5.9 &    26.6 \\  
\hline  
 {\bf M51:} \\   
 (0,0) &     17\,$10^3$ &   44.1 &    12.3 &    32.2 \\  
 (72,84) &     4.6\,$10^3$ &   16.7 &    $<2.3$ &    13 \\  
 (-84,-84) &     3.5\,$10^3$ &   15.4 &    2.1 &    13.4 \\  
\hline  
\end{tabular}
\end{center}
\end{table}

\subsection{FIR line fluxes from  ISO/LWS}
\label{sec-isolws_data}

The central area of M83 covering $\sim5'\times4'$ was mapped on a
fully-sampled grid of 61 positions with ISO/LWS. M51 was observed at
13 positions, mainly along a cut through the center and the two
prominent \CII\ lobes in the north-east and south-west seen in the KAO
map by \citet{nikola2001}.  The ISO/LWS line emission data was
uniformly processed by Brauher (2005, priv. com.) to derive line
fluxes in Wm$^{-2}$.
To convert to intensities, we use a LWS beam size of $80''$
($\Omega_{\rm LWS}=1.2\,10^{-7}$\,sr), the mean value published in the
latest LWS Handbook, and extended source corrections factors
\citep{gry2003}.  
%
Resulting ISO intensities of \CII\ (158\,$\mu$m, \NII\ (122$\,\mu$m),
and \OI\ (63$\,\mu$m) at the positions observed in \CI\ are listed in
Table\,\ref{tab-isodata}. 
Since the \OI\ (146$\,\mu$m) line was detected only at the center
positions (S.\,Lord, priv. comm.), we did not include it in the
present analysis.
%
%
%
%

\subsection{FIR continuum maps taken with IRAS}
\label{sec-iras_data}

To derive the far-infrared continuum at all selected positions, we
obtained high-resolution (HIRES) 60$\,\mu$m and 100$\,\mu$m IRAS maps
from the IPAC data center\footnote{A description of IRAS HiRes
  reduction is available at {\tt
    http://irsa.ipac.caltech.edu/IRASdocs/hires\_over.html}}.
Enhanced resolution images were created after 200 iterations using the
maximum correlation method \citep[MCM][]{aumann1990}. These data were
smoothed to an effective common circular beam of $80''$ and then
combined to create maps of the far-infrared flux. The FIR flux is
defined as in \citet{helou1988}: FIR $= 1.26\,10^{-14}
[2.58\,I(60\,\mu\rm{m})+I(100\,\mu\rm{m})]$ where FIR is in
W\,m$^{-2}$ and $I$ is in Janskys. FIR is a good estimate of the flux
contained between 42.5 and 122.5\,$\mu$m \citep{helou1988}.
Table\,\ref{tab-isodata} lists the FIR flux at the selected positions.
%
%
To derive the total infrared flux TIR, we follow the procedure
introduced by \citet{dale2001} who derived an analytical expression
for the ratio of total infrared flux TIR to the observed FIR flux,
i.e. the bolometric correction, as a function of the
$60\,\mu$m/100$\,\mu$m flux density ratio from modelling the infrared
SEDs of 69 normal galaxies
\footnote{ Note that TIR corresponds to the
  bolometric FIR dust continuum emission I(FIR) used in the PDR models
  of \citet{kaufman1999}.}
For M83 and for M51, the total infrared flux is a factor 2.3 larger
than FIR given the global $60\,\mu$m/100$\,\mu$m ratio of 0.43
(Table\,\ref{tab-the-sample}).

\section{Spectra and line ratios} 

The main aim of this work is to use the combined FIR ISO/LWS,
HIRES/IRAS, and the millimeter/submillimeter line data coherently and
to investigate, to what degree these give a consistent fit within the
framework of a simple model scenario such as PDR excitation.  We
therefore smoothed the $^{12}$CO maps (Table\,\ref{tab-data-sets}) to
the ISO/LWS angular resolution of $80''$ using Gaussian kernels.
Spectra of \CI, CO, and $^{13}$CO are displayed in
Figures\,\ref{fig-m83-spec},\ref{fig-m51-spec} and integrated
intensities in Table\,\ref{tab-intensities}. Ratios with \CI\ and
$^{13}$CO for which no maps exist, were corrected for beam filling
(Table\,\ref{tab-ratios}).



\subsection{M83}
\label{sec_m83_spectra}

The \CI\ lines are widest at the center position with 130\,kms$^{-1}$ 
 FWHM
and drop to about 30\,kms$^{-1}$ at the two outer positions. In
addition, peak line temperatures drop strongly, leading to a
pronounced drop of \CI\ integrated intensities and area integrated
\CI\ luminosities by factors of 10 to 18 at galacto-centric distances
of less than 2\,kpc (Table\,\ref{tab-intensities}).

 The CO 3--2 transition traces warm and dense gas since its upper
  level energy corresponds to $2.8\,J(J+1)=33.6\,$K and the critical
  density needed to thermalize this line is
  $4\,10^3\,J^3\sim1\,10^5\,$cm$^{-3}$, only weakly dependent on the
  kinetic temperature.  Trapping typically reduces the critical
  densities by up to an order of magnitude, depending on the optical
  depth of the lines.
  The CO 3--2/1--0 line ratio of integrated intensities
  (Table\,\ref{tab-ratios}) thus is a sensitive tracer of local
  densities for densities of less than $\sim10^5$\,cm$^{-3}$.  
Here, both line maps were smoothed to $80''$ resolution, beam filling
factors thus cancel out to first order.  The estimated calibration
error is 21\%. While the 2--1/1--0 ratio is $\sim1$ at all positions
indicating that the $J=2$ state is thermalized, the 3--2/1--0 ratio
drops slightly from 0.60 in the center to 0.46 at the spiral arm
positions. 
%
%
These ratios indicate that densities are lower than $1\,10^5$ needed
for thermalization of the $J=3$ level.

The $^{12}$CO 2--1 vs. $^{13}$CO 2--1 line ratio, as well as the
corresponding 1--0 ratio, trace the total column densities of the
$^{13}$CO line.  The ratios remain constant at $\sim10$ for 2--1 and
$\sim9$ for the 1--0 transition.

The above ratios found at the center position agree within the quoted
errors with the ratios presented in \citet{israel_baas2001}.

In M83, we find a significant drop of the CI vs.  $^{13}$CO 2--1 ratio
from about 4 in the center to $\sim1.4$ at the two outer positions.

\subsection{M51}
\label{sec_m51_spectra}

Again, the \CI\ line is widest at the center, $\sim100$\,kms$^{-1}$
 FWHM,
and drops to $\sim30$\,kms$^{-1}$ at the two outer positions. The
$^{13}$CO lines show similar line widths at all three positions.
%
Peak temperatures of \CI\ hardly drop between the center and the two
outer positions. Resulting integrated intensities and luminosities
drop by a factor of $\sim4$ only.

The observed CO 2--1/1--0 ratios lie between 0.6 and 0.8 for all three
positions and do not peak at the center. This indicates that not even
the 2--1 line is thermalized in M51. However, the CO 3--2/2--1 ratios
do not drop as would be expected but equal the CO 2--1/1--0 ratio or
even exceed them, while staying below 0.8. This is difficult to
explain with a single component model as we will show below.  The
center CO 3--2/1--0 ratio of 0.55 is in agreement with ratios
previously found with single-dish telescopes which range between
$0.5-0.8$ at beam sizes of $\sim14''$ to $24''$
\citep{matsushita1999,mauersberger1999,wielebinski1999}.
Interferometric observations at $\sim4''$ resolution tracing the dense
nuclear gas show a high 3--2/1--0 ratio of 1.9 \citet{matsushita2004}.

High central column densities are indicated in this work by the rather
low $^{12}$CO/$^{13}$CO 2--1 ratio of 4.6. In contrast, the outer
postions show ratios between 6 and 14.

We find CI vs.  $^{13}$CO 2--1 ratios of $\sim1$ at the center and at
($-84,-84$) while (72,84) exhibits a high ratio of 3.2.

\subsection{Comparison}

The gradients of \CI\ luminosities with galacto-centric distances are
strikingly different in M51 and M83. M83 is much more centrally
peaked, \CI\ luminosities drop by a factor of 18 at only 1.8\,kpc
distance in M83. In contrast, luminosities in M51 drop by only a
factor of 4 at galacto-centric distances which are more than a factor
of 3 larger, i.e. at 5.8\,kpc. However, the central \CI\ 
luminosities of M51 and M83 agree within 40\%.

The line widths observed at the outer positions of M51 and M83 are
typical for the disks of these two galaxies \citep{handa1990,gb1992}.
See Table\,2 of \citet{gb1993b} for a compilation of CO line widths
found in these and several other galaxies.

The CI vs.  $^{13}$CO 2--1 ratio in the Milky Way is often found to be
1 \citep[e.g.][]{keene1995} while \citet{israel2004,israel_baas2002}
find a strong variation of this ratio for 15 galactic nuclei. The \CI\ 
line is stronger than the $^{13}$CO 2--1 line for all but three galaxy
centers. The highest ratios are about 5. Here, we find a variation
between 1.3 and 4.

For galaxy centers, \citet{israel_baas2002} found that this ratio is
well correlated with the \CI\ luminosity covering a range of
160\,Kkms$^{-1}$ kpc$^2$ in the active nucleus of NGC3079 down to
$\sim1$\,Kkms$^{-1}$ kpc$^2$ in the quiescent center of Maffei\,2.
Here, we increase the range down to 0.11\,Kkms$^{-1}$ kpc$^2$
(Table\,\ref{tab-intensities}) at the same resolution of $10''$. In
contrast to the galaxy centers, the spiral arm positions observed here
do not show a systematic correlation between the \CI/$^{13}$CO line
ratio and \CI\ luminosity.

\begin{center}
\begin{table*}[h*]
\caption[]{\label{tab-intensities}
Integrated intensities in Kkms$^{-1}$ 
 (all columns but col.\,1, 2 and 4) at
the resolutions used for the spectra in Figs.\,\ref{fig-m83-spec},\ref{fig-m51-spec}.
Column 2 gives galacto-centric distances $R_{\rm gal}$ in kpc.
The calibration error is estimated to be $\sim15\%$.
 (The observed \CI\ intensities at the center positions agree within 20\% with
previous data presented in \citet{gerin_phillips2000} and \citet{petitpas_wilson1998}.)
Column 4 lists \CI\ luminosities in Kkms$^{-1}$ kpc$^2$ in brackets, i.e. the 
intensities integrated over the $10''$ JCMT beam. 
}
\begin{tabular}{lrrrrrrrrrr}
\hline 
\hline 
 ($\Delta \alpha$,$\Delta \delta$) & $R_{\rm gal}$ & \multicolumn{2}{c}{[CI]} & CO 1--0 & CO 2--1 & CO 3--2 & $^{13}$CO 1--0 & $^{13}$CO 2--1 \\  
  & & \multicolumn{2}{c}{{\rm HPBW=}$10''$} & $80''$ & $80''$ & $80''$ & $22''$ & $10''$ \\  
 (1) & (2) & (3) & (4) & (5) & (6) & (7) & (8) & (9) \\  
\hline 
{\bf M83:} \\  
 (0,0)   & 0  & 79.55  & (2)  & 50.63  & 47.03  & 30.15  & 19.36  & 21.33  \\  
 (-80,-72)   & 1.93  & 7.57  & (0.19)  & 18.17  & 18.06  & 8.35  & 4.6  & 5.1  \\  
 (89,38)   & 1.76  & 4.53  & (0.11)  & 17.03  & 15.93  & 7.84  & 4.28  & 3.27  \\  
\hline 
{\bf M51:} \\  
 (0,0)   & 0  & 16.36  & (2.78)  & 43.42  & 31.53  & 23.68  & 7.34  & 12.65  \\  
 (72,84)   & 5.4  & 4.59  & (0.78)  & 15.32  & 12.21  & 9.19  & 2.27  & 1.43  \\  
 (-84,-84)   & 5.77  & 4.03  & (0.69)  & 17.29  & 9.85  & 7.41  & 4.01  & 3.09  \\  
\hline 
\end{tabular}
\end{table*}
\end{center}

\begin{center}
\begin{table*}[h*]
\caption[]{\label{tab-ratios}
Line ratios of integrated intensities in Kkms$^{-1}$. 
The $1\sigma$ errors are $\sim21\%$. 
The CO 3--2/1--0 ratio was added to allow an easier comparison with the literature.
To compare all data at a common resolution of $80''$, we scaled the \CI\ and $^{13}$CO data, for which maps 
are not available, using the beam filling factors $\Phi_B^{80/10}$ and $\Phi_B^{80/21}$.
The factor $\Phi_B^{80/10}$ for \CI\ and $^{13}$CO 2--1 data was derived from the ratio of $^{12}$CO 2--1 integrated 
intensities at $80''$ and at $10''$ resolution. Likeweise, the factor $\Phi_B^{80/21}$ for $^{13}$CO 1--0 data was
derived from CO 1--0.
}
\begin{tabular}{cccccccccccccccccccccccccccccc}
\hline \hline
 {$\Delta\alpha,\Delta\delta$} & $\Phi_B^{80/10}$ & $\Phi_B^{80/21}$ & CO 3--2 & & CO 3--2 & & CO 2--1 & & CO 1--0 & & CO 2--1 & & \CI\ 1--0 & & \CI\ 1--0\\ 
 \cline{4-4} \cline{6-6} \cline{8-8} \cline{10-10} \cline{12-12} \cline{14-14} \cline{16-16} 
{$['','']$} & & & CO 1--0 & & CO 2--1 & & CO 1--0 & & $^{13}$CO 1--0 & & $^{13}$CO 2--1 & & $^{13}$CO 2--1 & & CO 3--2\\ 
\hline 
{\bf M83:} \\  
 (0,0)   & 0.22  & 0.28  & 0.6  & & 0.64  & & 0.93  & & 9.34  & & 10.02  & & 3.73  & & 0.58  \\  
 (-80,-72)   & 0.37  & 0.45  & 0.46  & & 0.46  & & 0.99  & & 8.77  & & 9.58  & & 1.48  & & 0.34  \\  
 (89,38)   & 0.48  & 0.47  & 0.46  & & 0.49  & & 0.93  & & 8.46  & & 10.14  & & 1.38  & & 0.28  \\  
\hline 
{\bf M51:} \\   
 (0,0)    & 0.56  & 0.68  & 0.55  & & 0.76  & & 0.73  & & 8.31  & & 4.62  & & 1.4  & & 0.4  \\  
 (72,84)    & 0.62  & 0.73  & 0.6  & & 0.76  & & 0.8  & & 9.27  & & 13.81  & & 3.19  & & 0.3  \\  
 (-84,-84)    & 0.52  & 0.72  & 0.43  & & 0.75  & & 0.57  & & 6.04  & & 6.09  & & 1.25  & & 0.27  \\  
\hline  
\end{tabular}
\end{table*}
\end{center}

\begin{table*}[th]
\caption[]{\label{tab-one-comp-model}
Results of the escape probability analysis using the four observed line ratios
$^{12}$CO 3--2/2--1, $^{12}$CO 2--1/1--0, $^{12}$CO/$^{13}$CO 1--0, and 2--1.
$N$(CO) is the total CO column density per $80''$ beam.  $M$ is the total mass integrated over
the beam and assuming an abundance ratio of [CO]/[H$_2$]=$8.5\,10^{-5}$ \citep{frerking1982}. 
$n_{\rm av}$ is the average density assuming that emission stems from a sphere with beam diameter.
 }
\begin{tabular}{crrrrrrrr}
\noalign{\smallskip} \hline \hline \noalign{\smallskip}
  $\Delta\alpha/\Delta\delta$ & $\chi_{\rm min}^2$ & 
  $n_{\rm loc}$ & $T_{\rm kin}$ & $N$(CO)$/\Delta$v & 
  $N$(CO) & $M$ & $n_{\rm av}$ \\
 $['','']$ & & 
  [cm$^{-3}$] & [K] & [$10^{16}$\,cm$^{-2}$/kms$^{-1}$] &
 [$10^{16}$\,cm$^{-2}$] & [$10^6\,$\msun] & [cm$^{-3}$] \\
\noalign{\smallskip} \hline \noalign{\smallskip}
 {\bf M83:} \\
 ($  0$,$  0$) &    1.7 &    3000. &   15.0 &    3.2 &  15.79 &  65.16 &   0.42 &  \\
 ($-80$,$-72$) &    3.5 &    3000. &   12.5 &    3.2 &   7.58 &  31.29 &   0.20 &  \\
 ($ 89$,$ 38$) &    2.1 &    3000. &   12.5 &    3.2 &   6.69 &  27.60 &   0.18 &  \\
\hline
 {\bf M51:} \\
 ($  0$,$  0$) &    3.6 &   30000. &    12.5 &     3.2 &   17.41 &  484.67 &    0.18 &  \\
 ($ 72$,$ 84$) &    4.0 &    3000. &    15.0 &     3.2 &    2.07 &   57.76 &    0.02 &  \\
 ($-84$,$-84$) &    7.4 &   10000. &    12.5 &     3.2 &    5.46 &  151.89 &    0.06 &  \\
\noalign{\smallskip} \hline \noalign{\smallskip}
\end{tabular}
\end{table*}


\section{Physical Conditions} 

\subsection{Simple homogeneous models}
\label{sec-lte}

As shown above, the observed CO line ratios cannot be explained with a
simple LTE analysis.  As a first step in order to estimate the kinetic
temperatures and local densities of the CO emitting gas, we present in
this section the results of slightly more realistic escape probability
radiative transfer calculations of homogeneous spherical clumps
\citep{stutzki_winnewisser1985}. We assumed a $^{12}$CO/$^{13}$CO
abundance ratio of 40 \citep{mauersberger_henkel1993}. As input, we
use the four ratios of integrated intensities
(Table\,\ref{tab-ratios}): $^{12}$CO 3--2/2--1, $^{12}$CO 2--1/1--0,
$^{12}$CO/$^{13}$CO 1--0, and $^{12}$CO/$^{13}$CO 2--1. We calculated
model intensities of the three transitions for column densities
$10^{14}<N$(CO)/$\Delta$v/(cm$^{-2}$/(Kkms$^{-1}$)$<10^{22}$, local
H$_2$ densities $10<n_{\rm loc}/$cm$^{-3}<40$, and kinetic
temperatures $10<T_{\rm kin}$/K$<40$. The modelled ratios were
compared with the observed ratios taking into account the
observational error of 21\% to derive the $\chi^2$.  The best fitting
$N$(CO)/$\Delta$v, $n$, and $T_{\rm kin}$ together with the
corresponding minimum reduced $\chi^2$ are listed in
Table\,\ref{tab-one-comp-model}.


Below, we first describe the results of the one-component fits for the
positions observed in M83 and M51.  Much more complete physical models
of the emitting regions are presented in the next section\,\ref{sec-pdr-analysis}.


\paragraph{M83:}

At $80''$ resolution, the $^{12}$CO line ratios do not vary
significantly between the center and the two bright spiral arm
positions observed here.  A $^{12}$CO 3--2/1--0 ratio of 0.5 indicates
an excitation temperature of $\sim10$\,K assuming optically thick
 thermalized
emission and simply using the detection equation. The escape
probability analysis leads to a similar result
(Table\,\ref{tab-one-comp-model}). At all three positions, the ratios
are well modelled by a one-component model with a rather low kinetic
temperature of only $12-15\,$K and a density of $n_{\rm
  loc}=3\,10^3$\,cm$^{-3}$.
This result does not exclude the existence of a warmer and denser gas
phase as would be traced by higher CO transitions or the 63\,$\mu$m
\OI\ line as discussed below. For the center, \citet{israel_baas2001}
have in fact deduced a warmer phase by including observations of the
CO 4--3 line in their radiative transfer analysis.

The $^{12}$CO and $^{13}$CO line ratios found in M83 are
characteristic for dynamically active or starburst regions in the
classification scheme of \citet{papadopoulos2004}. In this scheme,
extreme starbursts would show similar $^{12}$CO 2--1/1--0 and
3--2/1--0 ratios but much higher $^{12}$CO/$^{13}$CO ratios.

\paragraph{M51:}

In the escape probability analysis of CO and $^{13}$CO ratios in M51,
we discarded solutions leading to temperatures below 12.5\,K and
densities below $10^3$\,cm$^{-3}$ as unphysical. Since the CO $J=$3--2
is strong at all positions, densities and temperatures must be higher.

The CO 3--2/2--1 ratios are $\sim0.8$ at all positions in M51,
significantly higher than in M83. This indicates higher densities than
found in M83. Indeed, the escape probability analysis finds densities
between $3\,10^3$ and $3\,10^4$\,cm$^{-3}$.  The $^{12}$CO/$^{13}$CO
ratios agree with this solution for low temperatures of $\sim 12\,$K.
However, the observed CO 2--1/1--0 ratios are too low to agree with
this solution.
Neither the temperatures nor the densities are well constrained, and
minimum chi squared values are high. This shows the short coming of a
one-component model even when trying to model only the three lowest
rotational CO transitions. And it is in fact in agreement with the
finding of \citet{gb1993a} who used the lowest two transitions of
$^{12}$CO and $^{13}$CO and couldn't find a set of $T_{\rm kin}$ and
$n_{\rm loc}$ fitting simultaneously the line ratios for the arms and
for the central position.

\clearpage 
\newpage 
\begin{figure}[h]
\centering
\includegraphics[angle=0,width=8cm]{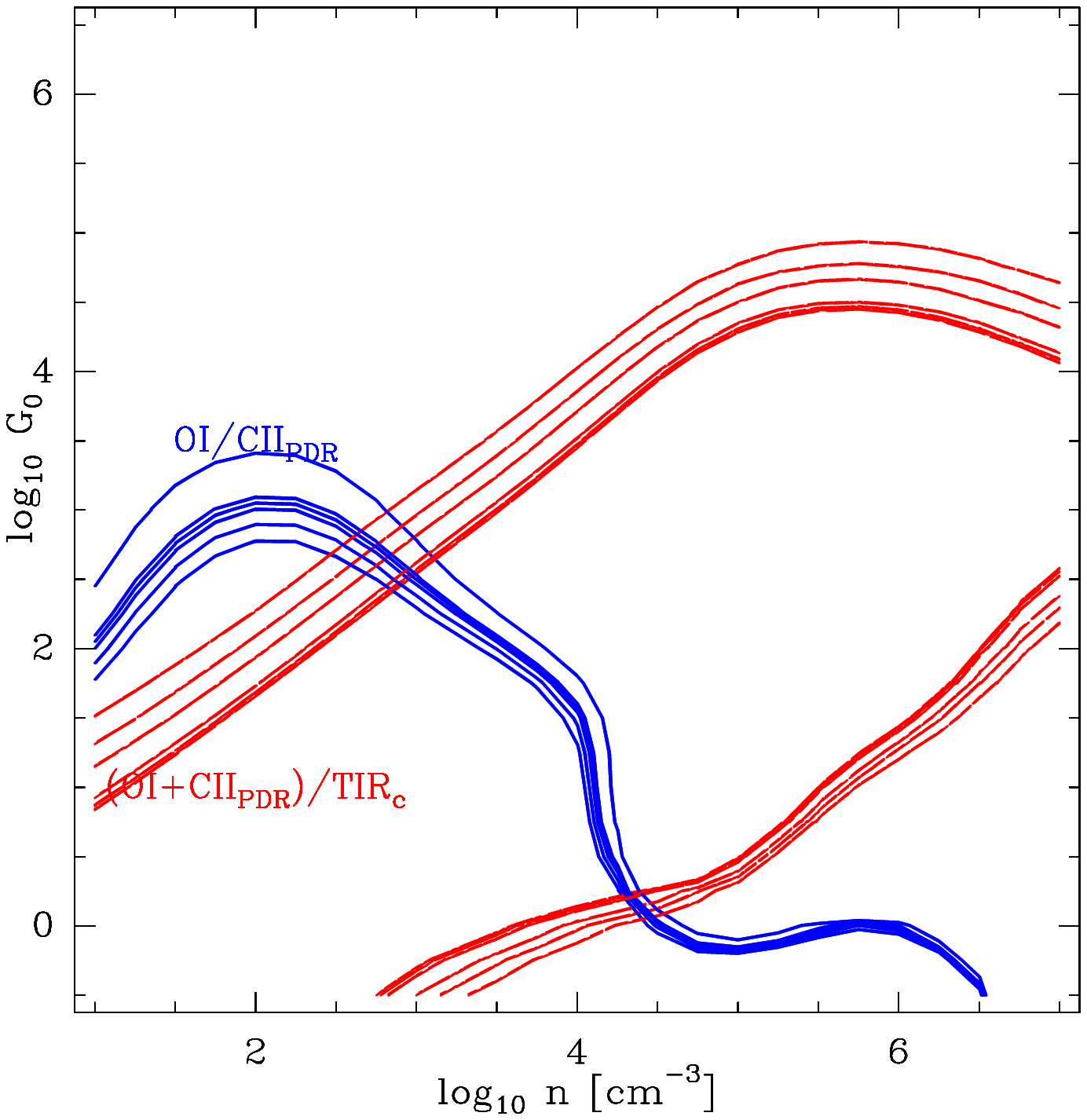} 
\caption{Comparison of the observed intensity ratios 
  \OI(63~\micron)/\CII$_{\rm PDR}$ and (\OI(63~\micron)+\CII$_{\rm
    PDR}$)/TIR$_c$ with PDR models \citep{kaufman1999} at the six
  positions in M83 and M51.
\label{fig_stand}}
\end{figure}

\begin{figure*}[h]
\centering
\includegraphics[angle=0,width=15cm]{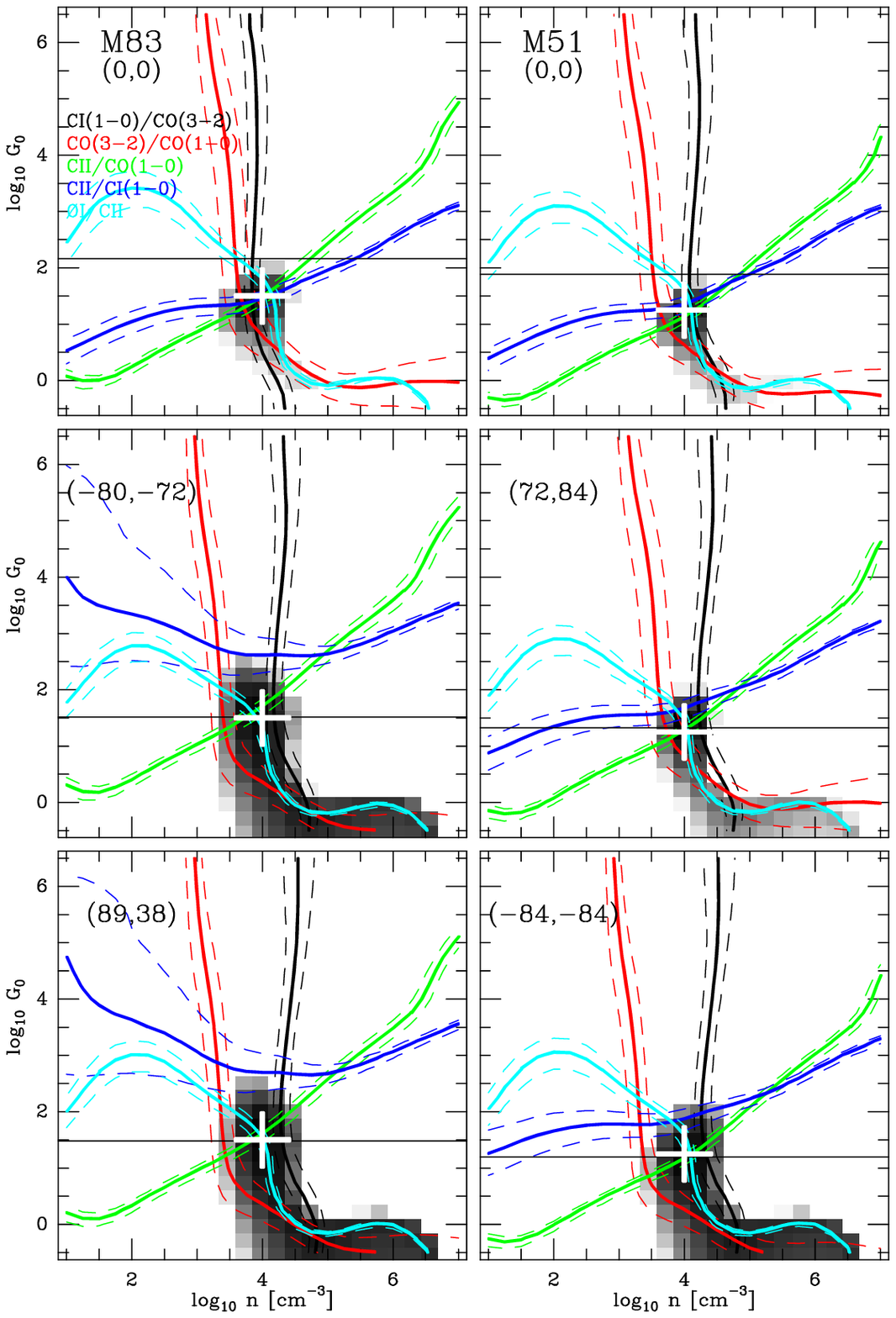} 
\caption{Comparison of the observed line intensity ratios 
  \CII$_{\rm PDR}$/\CI(1--0), \CI(1--0)/CO(3--2), CO(3--2)/CO(1--0),
  \CII$_{\rm PDR}$/CO(3--2) and \OI($63$~\micron)/\CII$_{\rm PDR}$.
  in M83 and M51 with PDR model calculations by \citet{kaufman1999}.
  The drawn contours correspond to the observed intensity ratios while
  the dashed contours are those for the 20\% uncertainty. The
  greyscale images show the reduced $\chi^2$ of the fit.  The position
  of the minimum reduced chi squared is marked by a white cross
  (Table\,\ref{tab_pdrfit}).  The horizontal black line shows the FUV
  flux $G_{\rm 0,obs}$ calculated from the observed TIR$_{\rm c}$ flux
  densities.
%
\label{fig_pdrplotm83m51}}
\end{figure*}

\subsection{PDR Analysis}
\label{sec-pdr-analysis}

To further constrain the physical conditions at the observed positions
in M83 and M51, we compare the observed line intensity ratios with the
results of the model for Photon Dominated Regions (PDRs) by
\citet{kaufman1999,tielens1985}.  The physical structure is
represented by a semi-infinite slab of constant density, which is
illuminated by FUV photons from one side. The model takes into account
the major heating and cooling processes and incorporates a detailed
chemical network.  Comparing the observed intensities with the
steady-state solutions of the model, allows for the determination of
the gas density of H nuclei, $n$, and the FUV flux, G$_0$, where G$_0$
is measured in units of the \citet{habing1968} value for the average
solar neighborhood FUV flux, $1.6\times10^{-3}$
ergs~cm$^{-2}$~s$^{-1}$. As has been pointed out in detail by
\citet{kaufman1999} in their Section 3.5.1 and several other authors,
the application of these models to extragalactic observations is not
straightforward since individual molecular clouds are not resolved in
single-dish observations and several phases of the ISM are therefore
observationally coexistent within each beam. The additional many
degrees of freedom in the parameter space for more complex models,
however, are ill constrained by the few observed, beam-averaged line
ratios. Hence, simplistic models, e.g. with only a single source
component, are used to at least derive average properties of the
complex sources. Nevertheless, it is possible to obtain some insight
into the spatial structure and the local excitation conditions, as we
will show.

%

\begin{table*}[htb]
\centering
\caption{Physical parameters at the observed positions M83 and M83, derived from
 fitting the observed intensity ratios to PDR models of \citet{kaufman1999}. 
 Column (2) gives the galacto-centric distance.
 Columns (3) and (4) list the fitted local densities and FUV fields together
 with the corresponding surface temperature (col.\,(5)) and the minimum
 chi squared (col.\,(6)). 
 The filling factor $\phi_{\rm UV}=G_{0,obs}/G_0$ reflects the ratio of TIR$_{\rm c}$ to
 the best fitting FUV field $G_0$. 
 $\phi_A^{\rm CI}$ is the area filling factor of the \CI\ emitting regions, i.e. the ratio of observed 
\CI\ intensity vs. that of the best fit model, corrected for beam and velocity filling (see text).
%
%
\label{tab_pdrfit}}
\begin{tabular}{lrrrrrrrrrrr} 
\hline
\hline
($\Delta\alpha$,$\Delta\delta$) & $R_{\rm gal}$ & $n$ & log(G$_0$) & T$_{\rm s}$ & 
$\chi^2$ & log(G$_{0,obs}$) & $\phi_{\rm UV}$ & $\phi_A^{\rm CI}$ \\ 
$['','']$ & [kpc] & [10$^4$cm$^{-3}$] &   & [K] &  & & & \\ 
(1) & (2) & (3) & (4) & (5) & (6) & (7) & (8) & (9) \\ 
\hline
  {\bf M83:} \\
 ($  0,  0$) &   0.00 &    4.0 &   1.50 &   76. &   2.8 &   2.16 &   4.57 &   0.019 \\ 
 ($-80,-72$) &   1.93 &    4.0 &   1.50 &   76. &  12.9 &   1.51 &   1.03 &   0.011 \\ 
 ($ 89, 38$) &   1.76 &    4.0 &   1.50 &   76. &  15.3 &   1.47 &   0.94 &   0.010 \\ 
  {\bf M51:} \\
 ($  0,  0$) &   0.00 &    4.0 &   1.25 &   66. &   3.2 &   1.88 &   4.31 &   0.013 \\ 
 ($ 72, 84$) &   5.40 &    4.0 &   1.25 &   66. &   5.2 &   1.32 &   1.17 &   0.016 \\ 
 ($-84,-84$) &   5.77 &    4.0 &   1.25 &   66. &  14.1 &   1.19 &   0.88 &   0.009 \\ 
\hline
\end{tabular}
\end{table*}

\subsubsection{\CII\ emission from the ionized and neutral medium}
\label{sec-cii-ion}

Carbon has a lower ionization potential (11.26\,eV) than hydrogen, so
that \CII\ emission arises not only from photon dominated regions, but
also from the ionized phases of the ISM and from the diffuse neutral
medium traced by \HI. 


Analyzing the Milky Way FIR line data obtained with FIRAS/COBE
\citep{fixsen1999}, \citet{petuchowski1993} argue that slightly more
than half the \CII\ emission in the Milky Way arises from PDRs, the
remainder from the extended low density warm ionized medium or diffuse
ionized medium (DIM), and an insignificant portion from the ordinary
cold neutral medium (CNM). 

\paragraph{\CII\ emission from the ionized medium.}
\label{sec_cii_di}

Here, we use the observed \NII(122$\mu$m) and \CII\ lines, to estimate
the fraction of \CII\ originating from the ionized medium.  However, a
thorough analysis would need more FIR emission line data from the
ionized medium, in particular the \NII(205$\mu$m) line, to
discriminate the relative importance of the different phases of the
ionized medium \citep{bennett1994}. Extragalactic observations of the
\NII(205$\mu$m) are however very rare to date \citep{petuchowski1994}.


The components of the ionized phase of the ISM which contribute to the
\CII\ emission are dense HII regions ($n_{\rm e}\ge~100$\,cm$^{-3}$)
and the diffuse ionized medium (DIM) ($n_e\sim~2$\,cm$^{-3}$)
\citep{heiles1994}.

%

The fraction of \CII\ stemming from \HII\ regions depends strongly on
the electron density of the ionized medium.  \citet{carral1994} showed
that upto 30\% of \CII\ stems from \HII\ regions when electron
densities exceed 100\,cm$^{-3}$. 
For dense HII regions ($n_{\rm e}\gg n_{\rm cr}$), model calculations
suggest 
$\CII/\NII\ = 0.28 ({\rm C/N})_{\rm dense}$ \citep{rubin1985} where
(C/N)$_{\rm dense}$ is the abundance ratio. The Galactic abundance
ratio found in dense \HII\ regions is 3.8 \citep{rubin1988,rubin1993}.
We thus expect to find an intensity ratio of
 \begin{equation} 
   \CII/\NII_{\rm ion,\HII} = 1.1.
 \label{eq-ciinii-hii}
 \end{equation}


Observations of radio free-free emission suggest that the DIM has
typical volume densities of n$_{\rm e}\sim2~{\rm cm}^{-3}$
\citep{mezger1978} and temperatures of T$_{\rm e} = 4000$~K
\citep{mueller1987}. Densities are thus considerably lower than the
critical densities of 
 310
\,cm$^{-3}$ for the \NII(122) line and 50\,cm$^{-3}$ for the \CII(158)
line
 \citep{genzel1991}.
In this limit, the total power emitted in the collisionally excited
fine-structure lines is simply the photon energy times the upward
collision rate, as given by \citet{heiles1994}. The resulting ratio of
intensities is: $\CII/\NII(122)\ = 3.05 {\rm (C/N)}_{\rm diff}$.
Galactic absorption line measurements of diffuse gas give (C/N)$_{\rm
  DIM} = 1.87$ \citep{sofia1997,meyer1997},
which implies an expected intensity ratio of
\begin{equation}
  (\CII/\NII(122))_{\rm ion,DIM}=5.7.
\label{eq-ciinii-dim}
\end{equation}


The preceeding section follows basically the arguments of
MKH01 and \citet{contursi2002}.


The metallicities, parametrized by the Oxygen abundances, have been
found to be only slightly supersolar in M83 and M51.
\citet{zaritsky1994} find (O/H) abundances of $1.5\,10^{-3}$ and
$1.9\,10^{-3}$ respectively, at 3\,kpc galacto-centric distance, from
visual spectra of \HII\ regions, which is about a factor 3 higher than
the solar metallicity of $0.46\,10^{-3}$ \citep{asplund2004}.
Abundance gradients with radius are found to be shallow in M83 and M51
\citep{zaritsky1994}. More recent observations with ISO/LWS and
modelling of the CCM10 \HII\ region of M51 by \citet{garnett2004}
indicate instead that the (O/H) abundances are about a factor of 2
less, i.e.  roughly solar. In addition, \citep{garnett1999} showed
that the (C/N) abundance ratio, which is of interest for the
\CII/\NII\ ratio discussed here, is independent of metallicity in both
normal and irregular galaxies.
%
We therefore use Galactic abundances to estimate the intensity ratios
in M83 and M51.  For the Milky Way, \citet{heiles1994} have estimated
that \NII\ originates predominantly from the DIM, contributing $\sim70$\%.
This was derived from the observed \NII(122)/\NII(205) ratio using
photo ionization models. However, the \NII(205) line has not yet been
observed in M83 and M51. We therefore use Eqs.\,\ref{eq-ciinii-dim}
and \ref{eq-ciinii-hii}, assuming that \NII\ stems solely from \HII\ 
regions, or alternatively, solely from the DIM.

Next, we can then derive the fraction of \CII\ emission originating from
PDRs:
\begin{eqnarray}
  \CII_{\rm PDR} & = & \CII_{\rm obs} - \Bigl(\frac{\CII}{\NII}\Bigr)_{\rm ion} \times \NII(122)_{\rm obs}
\label{eq-ciipdr}
\end{eqnarray}


The ratios of observed \CII\ versus \NII\ intensities vary between 3.6
and 7.4 at the six positions observed in M83 and M51
(Table\,\ref{tab-ciicorr}). \citet{garnett2004} found a higher ratio
of 11.8 at the CCM\,10 position in M51.

The ratios found in M83 and M51 lie at the low end of the ratios MKH01
found in the sample of 60 unresolved normal galaxies studied. They
show a mean ratio of 8 and a scatter between 4.3 and 29.
\citet{contursi2002} observed ratios of more than 7.7 in NGC6946 and
ratios of greater than 4 and 10 in NGC1313.  \citet{higdon2003} found
ratios between $\ge2.6$ and 20 in M33.

If the \NII\ emission originates only from the diffuse ionized medium, then
the major fraction of \CII\ arises from this phase, and only a small
fraction from PDRs (Table\,\ref{tab-ciicorr}). The observed \CII/\NII\ 
is $\le$5.8 at 3 positions, including the two nuclei, which would
indicate that no \CII\ emission at all arises from PDRs. This is
however unrealistic, since the emission of the \OI(63$\mu$m) line,
stemming from warm, dense PDRs, is strong compared to the \CII\ lines,
especially in the nuclei (Table\,\ref{tab-isodata}).  For this reason,
we discard this solution.

Assuming on the other hand, that the fraction of \CII\ from the
ionized medium and all \NII\ emission stem only from the dense \HII\ 
regions, then a fraction of only 15\% to 30\% of \CII\ originates from
this phase (Eq.\,\ref{eq-ciipdr}), while 70\% to 85\% of the observed
\CII\ emission then stems from PDRs. We prefer this solution and use
it in the PDR analysis discussed below. In an ISO/LWS study of
star-forming regions in M33, \citet{higdon2003} have recently used FIR
lines of the ionized medium, i.e.  \OIII(88$\mu$m), \OIII($52\mu$m)
and others to estimate the electron densities and other parameters of
the emitting gas, estimating that between 7\% and 47\% of \CII\ stems
from \HII\ regions, for their sample of positions.  They conclude that
the DIM is not needed to explain the observations.

                 
\begin{table}
\begin{center}
\caption{\label{tab-ciicorr} 
  FIR line ratios after correcting the observed \CII\ intensities for
  emission from the ionized medium. We study two cases: The emission
  of \CII\ from the ionized medium originates entirely {\bf a.} from the
  diffuse medium, and, {\bf b.}, that the emission arises entirely from
  dense \HII\ regions.
  Also shown is the intensity ratio $\epsilon$ of the two major
  cooling lines of PDRs (\OI(63)$+$\CII$_{\rm PDR}$) vs. the observed TIR
  continuum. }
\begin{tabular}[h]{crrrrrrrrr}
\hline \hline
 {$\Delta\alpha/\Delta\delta$} & \CII$_{\rm obs}$ & & \CII$_{\rm PDR}$ & & \OI(63) & $\epsilon$ \\
                                 \cline{2-2} \cline{4-4} \cline{6-6} 
                               & \NII(122)        & & \CII$_{\rm obs}$ & & \CII$_{\rm PDR}$ & [\%] \\
\hline
\multicolumn{5}{l}{{\bf a.} \NII\ only from the DIM:} \\
 {\bf M83:} \\
 $  0,  0$ &   5.8  & &  0.02  & & 57.41  &     0.12  \\ 
 $-80,-72$ &   7.3  & &  0.22  & &  3.18  &     0.22  \\ 
 $ 89, 38$ &   5.7 & &  0.00 & &  --  &     0.17 \\ 
 {\bf M51:} \\
 $  0,  0$ &   3.6 & &  0.00 & &  --  &     0.08 \\ 
 $ 72, 84$ &   7.3  & &  0.22  & &  3.59  &    0.16  \\ 
 $-84,-84$ &   7.4  & &  0.23  & &  3.78  &    0.21  \\ 
\hline
\multicolumn{5}{l}{{\bf b.} \NII\ only from \HII-regions:} \\
 {\bf M83:} \\
 $  0,  0$ &   5.8  & &  0.81  & &  1.31  &    0.21  \\ 
 $-80,-72$ &   7.3  & &  0.85  & &  0.83  &    0.36  \\ 
 $ 89, 38$ &   5.7  & &  0.81  & &  0.99  &    0.35  \\ 
 {\bf M51:} \\
 $  0,  0$ &   3.6  & &  0.69  & &  1.06  &    0.16  \\ 
 $ 72, 84$ &   7.3  & &  0.85  & &  0.91  &    0.25  \\ 
 $-84,-84$ &   7.4  & &  0.85  & &  1.02  &    0.33  \\ 
\hline
\end{tabular}
\end{center}
\end{table}


\paragraph{\CII\ emission from the diffuse neutral medium.}

The predicted \CII\ emission from the atomic gas has in general, for
many galactic nuclei, been found to be far too weak to account for the
observed \CII\ emission \citep{stacey1991} because the density is not
high enough to appreciably excite the \CII\ emission at the measured
\HI\ column densities. This view was confirmed by \citep{carral1994}
who conducted a detailed study of FIR cooling lines of NGC253 and
NGC3256.
%
Both in M51 and in M83, no large-scale correlation between \HI\ 
emission and that of \CII\ is seen, indicating again that \CII\ does not
trace the diffuse neutral medium \citep{nikola2001,crawford1985}.

\citet{nikola2001} used \HI\ column densities
\citep{tilanus_allen1991,rots1990} to derive the contribution to the
\CII\ emission in M51, assuming the same range of temperatures,
densities, and ionization fractions for the warm and cold neutral
medium (WNM, CNM) as have been found for the Milky Way.  They find
that the contribution of the WNM is negligible for most of the M51
disk except the northwest, which was not studied here.  The
contribution of the CNM is estimated to be less than 10\%$-$20\% in
all regions but the northwest.


We have thus not corrected the \CII\ emission for a possible
contribution from the diffuse neutral medium.

\subsubsection{The infrared continuum and the FUV field \label{sec_firfuvcomp}}

The stellar FUV photons heat the molecular gas and dust which
subsequently cools via the FIR dust continuum and, with a fraction of
less than $\sim1$\% \citep{stacey1991},MKH01 via \CII, \OI(63),
and other cooling lines. To the extent that filling factors are 1 and
other heating mechanisms like cosmic ray heating can be neglected, the
observed TIR continuum intensity should equal the modelled FUV field.
This is also expected if a constant fraction of FUV photons escape
without impinging on cloud surfaces.

The PDR model of \citet{kaufman1999} assumes a semi-infinite slab
illuminated from one side only. For the extragalactic observations
described here, we however have several PDRs within one beam and the
clouds are illuminated from all sides. Hence, the optically thin total
IR intensity stems from the near and far sides of clouds.  Here, this
is taken into account by dividing the observed TIR by 2
\citep[][]{kaufman1999}:
 ${\rm TIR}_{\rm c} = \rm{TIR}/2 = 2.3\times\rm{FIR}/2$ (cf.
    Sec.\ref{sec-iras_data}).
While this correction holds exactly only for
finite plane parallel slabs illuminated from both sides, it is a good
first approximation. The corrected TIR can then be used to derive the
corresponding FUV intensity via
$G_{0,{\rm obs}}$ =
TIR$_c$$4\pi/(2\times1.6\,10^{-3})$\,ergs~cm$^{-2}$~s$^{-1}$. 
 Following the arguments of \citet{kaufman1999}, the additional
  factor 2 takes into account equal heating of the grains by photons
  outside the FUV band, i.e. by photons of $h\nu<6\,$eV.  
We find a variation by one order of magnitude,
%
 $15<G_{0,{\rm obs}}<144$ 
%
%
(Table\,\ref{tab_pdrfit}, Fig.~\ref{fig_pdrplotm83m51}).
%
%
%


\subsubsection{First estimates of FUV field and density}

The two intensity ratios \OI(63)/\CII$_{\rm PDR}$ and
(\OI(63)$+$\CII$_{\rm PDR}$)/TIR$_c$, of the two major PDR cooling
lines and the continuum, have been used extensively to derive the
density and FUV field of the emitting regions
(e.g. MKH01). Since \OI(63) and \CII\ are the dominant
coolants, the latter ratio is a good measure of the photoelectric
heating efficiency $\epsilon$ \citep[e.g.][]{kaufman1999}. The former
ratio measures the relative importance of \CII\ vs \OI(63) cooling.
For high FUV fields and high densities, the ratio becomes larger than
one \citep{kaufman1999}.

In M83 and M51, the intensity ratio \OI(63)/\CII$_{\rm PDR}$ varies
only slightly between 0.8 and 1.3 (Table\,\ref{tab-ciicorr}b).  The
heating efficiency varies between $\sim0.25$ and 0.36\% at the
outskirt positions while it drops to below 0.21\% in the centers.

The values which we find in M83 and M51 lie within the range covered
by MKH01, who find heating efficiencies ranging between $\sim0.3$\%
and $\sim0.05$\% for the 60 galaxies studied, while the \OI/\CII\ 
ratios range between 0.3 and $\sim10$.  Though the scatter is large,
the heating efficiency tends to be high $>0.15\%$ for \OI/\CII\ ratios
of less than 2.  The \OI(63)/\CII$_{\rm PDR}$ ratios found in M83 and
M51 agree roughly with the average value found by MKH01, while the
heating efficiencies in M83 and M51 span the average value of MKH01
upto the highest efficiencies found by them.

MKH01 corrected the observed \CII\ emission by roughly 50\% when
taking into account the contribution from the ionized medium, based on
the Milky Way results. Here, we corrected by only 15\% to 30\%
 (Table\,\ref{tab-ciicorr}b).
This uncertainty in how best to correct the \CII\ fluxes, needs to be
considered when comparing the derived heating efficiencies and
\OI/\CII\ ratios.


The small scatter of the observed two ratios at the 6 positions in M83
and M51 indicates that the emitting gas has similar physical
properties.  Comparison with the results of the Kaufman PDR model
shows that two solutions exist (Figure\,\ref{fig_stand}).  The data
can be reproduced either by high FUV fields at low densities or by low
FUV fields and high densities.  The high-$G_0$ solution indicates
$2.5\le{\rm log}(n/{\rm cm}^{-3})\le3.2$ and $2.4\le{\rm log}G_0\le3$.
As we will show, the low-$G_0$ solution is less plausible. It
indicates rather high densities of $4.3\le{\rm log}(n/{\rm
  cm}^{-3})\le4.5$ and low FUV fields of $0.1\le{\rm log}G_0\le0.5$.
In this case, the observed \CII\ intensities are more than three
orders of magnitude larger than the modelled intensities which would
indicate that many PDR slabs along the lines of sight. Since the
optical depth of the \CII\ line in the line centers is expected to be
about one \citep{kaufman1999}, this scenario is discarded. This
argument also holds when velocity filling is taken into account, since
the velocity filling factor is $<40$ at all positions, as discussed
below.  We note that \citet{higdon2003} in their analysis of ISO/LWS
data of M33, also discussed the two possible PDR solutions, and,
following a different line of reasoning, also preferred the high-$G_0$
solution.

The $n,G_0$ values we find for the high-$G_0$ solution, agrees with
the average values found by MKH01 who also exclude the low-$G_0$
solution. Their sample of 60 unresolved galaxies covers a slightly
larger range of values: $2<{\rm log}(n/{\rm cm}^{-3})<4.25$ and
$2.5<{\rm log}G_0<5$. Our values also agree with the average value
found in NGC\,6946 by \citet{contursi2002}.

\subsubsection{PDR model fitting of five line ratios}

In order to determine with greater confidence the densities and UV
fluxes which can explain the intensity ratios, we have performed a
$\chi^2$ fitting of the observed 5 line intensity ratios \CII$_{\rm
  PDR}$/\CI(1--0), \CI(1--0)/CO(3--2), CO(3--2)/CO(1--0), \CII/CO(3--2),
and, \OI(63)/\CII$_{\rm PDR}$ relative to the predictions of the PDR
model by \citet{kaufman1999}:
$$\chi^2 = \frac{1}{4-2}\sum_{i=1}^{5} \Bigl( \frac{R^{\rm
    obs}_i-R^{\rm mod}_i}{\sigma_i}\Bigr)^2$$
with the observed and
modelled ratios $R^{\rm obs}_i$ and $R^{\rm mod}_i$, and the error
$\sigma_i$ which is assumed to be 20\% for all ratios\footnote{There
  are two degrees of freedom since we use four independent line
  ratios to fit two parameters.}.
%
%
In Fig.~\ref{fig_pdrplotm83m51} each panel shows the observed line
intensity ratios and the calculated reduced $\chi^2$ in greyscale for
each observed position in M83 and M51.  Table~\ref{tab_pdrfit}
summarizes the local density ($n$) and FUV radiation field ($G_0$) at
the position of minimum reduced $\chi^2$ corresponding to the best fit
models thus identified. 
 The \CI(1--0)/CO(3--2) and the CO(3--2)/(1--0) ratios are excellent
tracers of the local densities, almost independent of $G_0$ for
$G_0>10$.
The sensitivity of \CII\ on the FUV field is reflected by the line
ratios which include \CII. The combination of all these ratios
therefore allows, in principle, to deduce $n$ and $G_0$.


\paragraph{Densities.} 
Though the quality of the fit varies strongly between the six
positions, the best fitting FUV field and density are almost
identical: $n=10^4$\,${\rm cm}^{-3}$ and $18<G_0<32$
(Table\,\ref{tab_pdrfit}).  Model surface temperatures are
$\sim70$\,K. This is slightly lower than the upper energy level of
\CII(158), $E_{\rm up}/k=92$\,K, and much lower than the corresponding
level of \OI(63), $E_{\rm up}/k=228$\,K. Since the critical densities
of the two \OI\ lines for collisions with H are also high,
$>10^5$\,cm$^{-3}$, the \OI\ lines are subthermally excited.
Nevertheless, gas cooling is dominated by \CII\ and \OI, contributions
from H$_2$ or Si, covered by ISO/SWS, are negligible at these low
temperatures.  The best fitting density agrees within a factor of 3
with the local densities derived above via simple radiative transfer
analysis, using only the $^{12}$CO and $^{13}$CO line ratios
(Table\,\ref{tab-one-comp-model}).

\paragraph{UV filling factor.} 
At the four spiral arm positions, the total infrared continuum agrees
perfectly with the best fitting FUV field (Table\,\ref{tab_pdrfit}),
which is an independent confirmation of the validity of the PDR
analysis using the five line ratios. The active center regions of M83
and M51 show filling factors of $\sim4$, indicating that other sources
than PDRs heat the dust leading to the high observed TIR intensities,
e.g.  massive protostars.

\paragraph{Quality of the fits.}  The minimum reduced chi squared of
the 4 independent ratios lie between 3 and 15 at all positions
(Table\,\ref{tab_pdrfit}). While the two nuclei and the NE-spiral arm
position of M51 show $\chi^2$ of better than 5, the other three spiral
arm positions show rather poor $\chi^2$ values of 13--15. 

Inspection of Figure\,\ref{fig_pdrplotm83m51} shows that, at the
latter three positions, the \CII$_{\rm PDR}$/\CI\ ratio indicates
higher FUV fields than the best fitting solution.  In addition,
Figure\,\ref{fig_pdrplotm83m51} also shows that the CO 3--2/1--0 line
ratios indicates systematically lower densities than the observed
\CI/CO 3--2 ratios.  This holds to varying degrees for all positions
and for $G_0>10$. By assuming \CI\ intensities which are a factor 2
higher, the quality of the fit is considerably improved, while the
best fitting $G_0$ and density stay constant at all positions. At
($-80$,$-72$) in M83 for example, the $\chi^2$ is improved from 12.9
to 3.4. This indicates that the beam filling factors of the \CI\ 
emission, derived from the CO 2--1 data (Table\,\ref{tab-ratios}), are
too small, i.e. \CI\ is more extended than CO.

We also conclude that our results are consistent with the assumption
that only the dense ionized medium contributes to the \CII\ emission.
There is no need for an extended diffuse component.


Overall, the observed five ratios cannot be well fitted with a single
plane-parallel PDR model of constant density at any of the positions.
This is not surprising as the excitation requirements of the various
tracers collected here vary widely. Especially, the critical densities
vary between $5\,10^2$\,cm$^{-3}$ for the lower \CI\ transition and
$5\,10^5$\,cm$^{-3}$ for the \OI\ 63$\mu$m transition. Any density
gradients in the emitting medium may thus lead to the above
discrepancies with a single PDR model depending also on the chemical
and temperature structure. In addition, the large difference between
local densities derived here and beam averaged densities of more than
three orders of magnitude (Table\,\ref{tab-one-comp-model}) shows that
the emitting volume must be filled with very small but dense
structures.  From many Galactic observations, it is expected that
these structure show a spectrum of masses, adding to the complexity
ignored here.

\paragraph{Absolute intensities.} 

When comparing the observed intensities with the model results, the
velocity filling has also to be taken into account.  The observed \CI\ 
line widths $\Delta{\rm v}_{\rm obs}$ range between 30 and
130\,kms$^{-1}$ FWHM (cf.\,Sec.\ref{sec_m83_spectra} and
\ref{sec_m51_spectra}) measuring the dispersion of clouds within the
beam for these extragalactic observations. On the other hand, the
microturbulent velocity dispersion $\delta_v$ of the gas of one PDR
model is set to 1.5\,kms$^{-1}$ \citep{kaufman1999}, corresponding to
a Gaussian FWHM $\Delta{\rm v}_{\rm mod}$ of 3.5\,kms$^{-1}$ as is
typical for individual Galactic clouds. To calculate the \CI\ area
filling factors (Table\,\ref{tab_pdrfit}), we divided the observed
\CI\ intensities by the predicted \CI\ intensity from the best fitting
model, corrected for beam (Table\,\ref{tab-ratios}) and velocity
filling, viz.,
$$\Phi_A^{\rm CI}=\frac{({{\rm [CI]}_{\rm obs}}\times\Phi_B^{80/10})}
{{\rm [CI]}_{\rm mod}}\times\frac{\Delta{\rm v}_{\rm mod}}{\Delta{\rm
    v}_{\rm obs}}.$$

\CI\ emission fills only a few percent of the $80''$ beam, the derived
area filling factors vary between $(1-2)\,10^{-2}$
(Table\,\ref{tab_pdrfit}), consistent with the low volume filling
factors described above.

\section{Summary} 

We have studied all major submillimeter and far infrared cooling lines
together with the dust total infrared continuum at the center
positions of the two galaxies M83 and M51 and at four spiral arm
positions. 
  
  We observed \CI\ 1--0 at the six positions at $10''$ resolution.
  Complementary \CII, \OI(63), and \NII(122) data were obtained from
  ISO/LWS at $80''$ resolution. CO maps of the lowest three
  transitions were obtained from the literature and smoothed to the
  ISO/LWS resolution. We also obtained pointed $^{13}$CO 1--0 and 2--1
  data at all positions. In order to allow a comparison of all these
  data, the \CI\ and $^{13}$CO data were scaled with beam filling
  factors derived from the $^{12}$CO data. For completeness, we also
  obtained the total far-infrared continuum intensities from
  HIRES/IRAS $60\,\mu$m and $100\,\mu$m data.

\begin{itemize}
  
\item Integrated intensities peak in the two centers. However, M83 is
  much more centrally peaked than M51 as seen in CO and \CI\ as is
  already seen in the spectra
  (Figs.\,\ref{fig-m83-spec},\ref{fig-m51-spec}).  This is seen even
  more drastically in the drop of \CI\ luminosities with
  galacto-centric distance. In M83, luminosities drop by more than one
  order of magnitude over $\sim2$\,kpc, while in M51, they drop by
  only a factor of $\sim$4 over $\sim6$\,kpc. Obviously, this analysis
  should be refined by more observations of \CI\ at different radii.
  
\item The $^{12}$CO 3--2/1--0 line ratios lie below 0.6 at all
  positions, indicating subthermal excitation of the 3--2 line, i.e.
  densities are less than $10^5$\,cm$^{-3}$. This is confirmed by more
  detailed analysis using escape probability and PDR models. The
  former homogeneous models indicate densities between $3\,10^3$ and
  $3\,10^4$\,cm$^{-3}$ from CO and $^{13}$CO line ratios.
  
\item We estimated the fraction of observed \CII\ emission originating
  from the ionized medium by using the \NII\ 122$\,\mu$m data which
  solely traces this medium. In the absence of additional data, which
  would allow to seperate the contributions from the different phases
  of the ionized medium, we argue that dense \HII\ regions are the
  primary source of \NII\ emission. These emit 15\% to 30\% of the
  observed \CII\ emission while the remainder stems from PDRs.
  
\item The gas heating efficiency was calculated from the ratio of the
  two major gas coolants \OI\ at 63\,$\mu$m and \CII$_{\rm PDR}$ at
  158\,$\mu$m versus the total infrared intensity. The efficiency is
  low in the centers showing ratios of (1.6--2.1)\,$10^{-3}$, while
  the outer positions show higher ratios of (2.5--3.6)\,$10^{-3}$. The
  latter efficiencies lie at the high end of efficiencies observed by
  MKH01 in a sample of 60 unresolved normal galaxies. 
  
\item We fitted the observed line intensity ratios \CII$_{\rm
    PDR}$/\CI(1--0), \CI(1--0)/CO(3--2), CO(3--2)/CO(1--0),
  \CII/CO(3--2), and, \OI(63)/\CII$_{\rm PDR}$ to the predictions of
  the PDR model of \citet{kaufman1999}. The best fits yield densities
  of $10^4$\,cm$^{-3}$ and FUV fields of $\sim\,G_0=20-30$ times the
  average interstellar, almost constant at the six positions studied
  here at $80''$ resolution. This finding may be a selection effect,
  since we selected the four outer positions for their high star
  forming activity. More observations of other less prominent
  positions are needed to study variation of physical parameters over
  the galaxy surfaces.
  
\item Filling factors vary significantly between the center positions
  and the outer positions in both galaxies. The filling factor
  $\phi_{\rm UV}$ derived from the FUV field calculated from the
  observed total infrared intensity versus the fitted $G_0$ lies at 1
  at the spiral arm positions, with a scatter of less than 20\%. In
  contrast, the nuclei show ratios of $\sim4-5$. The density contrasts
  within the emitting gas must be high, given the local densities of
  $10^4$\,cm$^{-3}$ and the average densities, derived from CO column
  densities, of less than 1\,cm$^{-3}$, leading to very low volume
  filling factors. In accordance, the area filling factors of the \CI\ 
  emission are less than 2\% at all positions.
  
\item The reduced chi squared lie between 3 and 5 at three positions,
  including the two nuclei, while the fits are worse at the three
  other positions, which show $\chi^2$ = 13--15. It is shown that the
  fits can be significantly improved by assuming that \CI\ emission is
  more extended.
  

\end{itemize}

While the present analysis led to the results listed above, it needs
to be refined with new data and improved modelling. The assumption
that all gas tracers have the same filling factors may not be
justified. We find indications that \CI\ is more extended than CO
2--1. Clearly, maps of \CI\ at $10''$ resolution filling the ISO/LWS
$80''$ beam would be needed to verify this conclusion.  A related
question is the nature of the interclump medium and whether it can be
ignored when interpreting the submm/FIR emission. The fraction of
\CII\ emission from the different phases of the ionized medium has
been addressed here by assuming that all emission stems from the dense
ionized medium. While this is consistent with the PDR modelling, it
probably is an oversimplification. Observations of the \NII\ line at
$205\,\mu$m would allow to refine this analysis. Such observations
will become possible with a new generation of submm telescopes and
receivers operating at high altitudes. In this regard, modelling neads
to include the ionized medium.  Improved PDR models need to consider a
distribution of clouds within each beam, following mass and size
distribution laws. 


%

\begin{acknowledgements}
  We thank Steve Lord for valuable discussions.
  We would like to thank the JCMT and the IRAM 30m staff for providing
  excellent support during several long runs at the Mauna Kea and the
  Pico Veleta.
%
  We are greatful to Michael Dumke for providing us with HHT CO 3--2
  and 4--3 M51 data and Lucian Crosthwaite to provide NRAO CO 1--0 and 2--1
  M83 data. 
%
%
  And we thank Jim Brauher, Steve Lord, and Alessandra Contursi
  for providing us with the ISO/LWS line fluxes of both galaxies.
%
%
  The James Clerk Maxwell Telescope is operated by the Joint Astronomy
  Centre on behalf of the Particle Physics and Astronomy Research
  Council of the United Kingdom, the Netherlands Organisation for
  Scientific Research, and the National Research Council of Canada.
  This work has benefited from research funding from the European
  Community's Sixth Framework Programme.
  We made use of the NASA IPAC/IRAS/HiRes data reduction facilities.
\end{acknowledgements}

\bibliographystyle{aa}
\bibliography{aamnem99,p_galaxies_bib} 

\end{document}